\documentclass[5p,number,sort&compress,round,times,fleqn]{elsarticle}
\usepackage[numbers]{natbib} 
\setcitestyle{square}

\makeatletter
\def\ps@pprintTitle{%
  \let\@oddhead\@empty
  \let\@evenhead\@empty
  \let\@evenfoot\@oddfoot
}
\makeatother

\usepackage{graphicx} 
\usepackage{float} 
\usepackage{color} 
\usepackage{amsmath} 
\usepackage{amssymb} 
\usepackage{mathtools} 
\usepackage[only,llbracket,rrbracket]{stmaryrd} 
\usepackage[misc]{ifsym} 
\usepackage{bm} 
\usepackage{setspace} 
\usepackage{xspace} 
\usepackage{hyphenat} 
\usepackage{comment} 
\usepackage{array} 
\usepackage{makecell} 
\usepackage{tabularx}
\usepackage[flushmargin]{footmisc} 
\usepackage[colorlinks,pdfpagelabels,bookmarksopen,bookmarksnumbered,citecolor=blue,urlcolor=blue,linkcolor=blue]{hyperref} 

\floatstyle{ruled}
\newfloat{Fbox}{thp}{lop}
\floatname{Fbox}{Box}

\newdefinition{rmk}{Remark}
\newproof{pf}{Proof}
\newproof{pot}{Proof of Theorem \ref{thm2}}

\newcommand{\bs}[1]{\mathbf{#1}} 
\newcommand{\bms}[1]{\bm{#1}} 
\newcolumntype{C}[1]{>{\centering\arraybackslash}p{#1}}

\newcommand{\old}[1]{\iffalse#1\fi}

\begin{document}

\title{Multiresonant Layered Acoustic Metamaterial (MLAM) solution for broadband low-frequency noise attenuation through double-peak sound transmission loss response}

\author[add1,add2]{D. Roca}
\author[add1,add2]{J. Cante}
\author[add1,add3]{O. Lloberas-Valls}
\author[add4,add3]{T. P\`{a}mies}
\author[add1,add3]{J. Oliver\corref{cor}}\ead{oliver@cimne.upc.edu}

\cortext[cor]{Corresponding author}

\address[add1]{
	Centre Internacional de M\`{e}todes Num\`{e}rics en Enginyeria (CIMNE)\\
	Campus Nord UPC, Ed. C1, C/ Gran Capit\`{a} s/n, 08034 Barcelona (Spain)
}
\address[add4]{
    Laboratori d'Enginyeria Acustica i Mec\`{a}nica (LEAM)\\
    Campus Terrassa UPC, C/ Colom 11, 08222 Terrassa (Spain)
}
\address[add2]{
	Universitat Polit\`{e}cnica de Catalunya (UPC), Escola Superior d'Enginyeries Industrial Aeroespacial i Audiovisuals de Terrassa (ESEIAAT)\\
	Campus Terrassa UPC, C/ Colom 11, 08222 Terrassa (Spain)
}
\address[add3]{
	Universitat Polit\`{e}cnica de Catalunya (UPC), Escola T\`{e}cnica Superior d'Enginyers de Camins, Canals i Ports de Barcelona (ETSECCPB)\\
	Campus Nord UPC, Ed. C1, C/ Jordi Girona 1, 08034 Barcelona (Spain)
}

\begin{abstract}
	The problem of noise control and attenuation is of interest in a broad range of applications, especially in the low-frequency range, below 1000~Hz. Acoustic metamaterials allow us to tackle this problem with solutions that do not necessarily rely on high amounts of mass, however most of them still present two major challenges: they rely on complex structures making them difficult to manufacture, and their attenuating capabilities are limited to narrow frequency bandwidths. Here we propose the Multiresonant Layered Acoustic Metamaterial (MLAM) concept as a novel kind of acoustic metamaterial based on coupled resonances mechanisms. Their main advantages hinge on providing enhanced sound attenuation capabilities in terms of a double-peak sound transmission loss response by means of a layered configuration suitable for large scale manufacturing.
\end{abstract}


\maketitle

\section{Motivation}
\label{sec:1}	
\begin{figure}
	\includegraphics{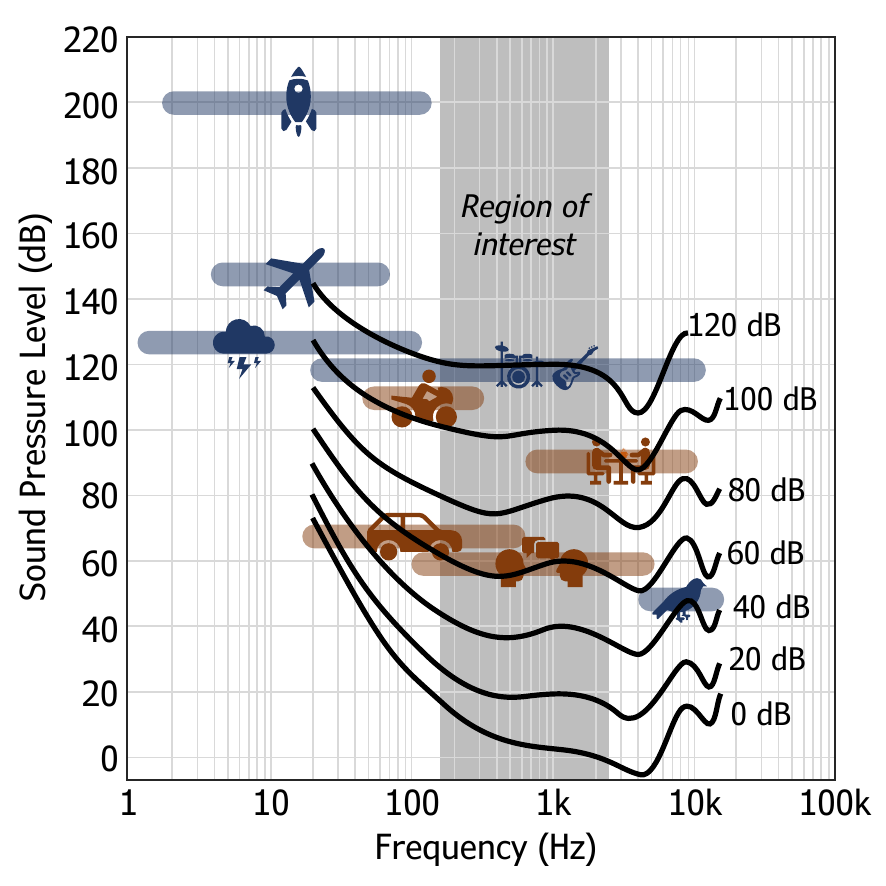}
 	\caption{Sound pressure level and frequency range for common noise sources.~Each solid black curve is associated to a decibel level at which each sound source is perceived by human hearing at different frequencies.~As an example, traffic noise at 70~dB is perceived around this same level at frequencies in the range of 200-2000~Hz, but below 100~Hz the perceived intensity rapidly decreases, becoming practically inaudible below 30~Hz.~The shaded area indicates the frequency range of interest, where most common noise sources can be found, and their perceived intensities are higher.\label{Fig1}}
\end{figure}

For years, the notion of metamaterials has attracted the interest among the scientific community for their potential range of applications.~Their attractiveness resides on the ability to exhibit customized behavior, specifically in the manipulation of waves, by engineering their internal structure.~While the concept was originated in the field of electromagnetism, with metamaterials that showed both negative permittivity and permeability \cite{Veselago1968}, the idea rapidly extended to other areas.~In regards of the present work, the focus is on the manipulation of acoustic waves, aiming at designing a metamaterial with enhanced sound attenuation capabilities in the low-frequency region, i.e. between 100 and 1000~Hz, where most common noise sources occur (see Fig.~\ref{Fig1}).\\
\indent The first notions of acoustic metamaterials emerged two decades ago, when Ping Sheng and co-workers demonstrated the concept of local resonance with a metamaterial composed of rubber-coated lead spheres, acting as internal resonators, embedded in an epoxy matrix \cite{Liu2000}.~Several other works followed, aiming at exploiting the local resonance phenomena involved in the production of frequency band gaps for improved acoustic attenuation in selected frequency regions.~Beyond the early designs based on the combination of rubber-coated inclusions in a polymer matrix \cite{Sheng2003,Calius2009,Wester2009}, examples including membranes as internal resonators \cite{Yang2008,Khanolkar2015,Hiraiwa2016}, as well as periodic arrangements of pillars on a plate \cite{Badreddine2012,Badreddine2012b,Bilal2017}, have been explored, among others.~Non-periodic acoustic metamaterial configurations based on irregularities and random structures have also been studied in the past years \cite{Rupin2014,Celli2019}.\\
\indent The main challenges faced by most of the proposed resonance-based acoustic metamaterials so far are: (1) the frequency bandgaps they typically produce are narrowband, and (2) they cannot be easily manufactured.~The former issue is related to the local resonance phenomenon itself so it mainly affects the low frequency range (for higher frequencies, Bragg scattering effects typically produce wider bandgaps).~In order to produce wider frequency bandgaps, one needs to increase the internal resonating mass in the metamaterial, which makes them lose their lightweight capabilities that makes them competitive against other classical approaches.~Several works found in the literature have tried to overcome this issue, for instance, by employing multiple resonators \cite{Xiao2012,Zhao2016,Zhou2020} or combining local resonance with other damping mechanisms \cite{Hussein2013,Manimala2014,Krushynska2016,VanBelle2019}.~However, the problem has not been entirely addressed.~Multiple resonators typically lead to a collection of separated narrowband frequency bandgaps.~As for the enhancements obtained through the introduction of dissipation phenomena, they are somewhat limited to very specific damping ranges (which may not be easily obtained, in practice, with conventional materials).\\
\indent As for manufacturability, most of the proposed acoustic metamaterials resort to combine different material components, whose assembly presents a challenge for their practical implementation, for instance, in a mass production chain.~Furthermore, ideas based on a single material component, which typically provide a solution to this problem, rely on additive manufacturing technologies due to their topological and geometrical complexity and characteristics \cite{Claeys2016,Leblanc2017,McGee2019}.~This kind of technology, while presenting itself as a potential promising manufacturing option in the future, still faces many challenges nowadays, especially in terms of processing times and dimensional tolerances (see Section~\ref{sec:4.2} for a further insight on that), which are of utmost importance in order to produce effective acoustic metamaterials in a practical manner.\\
\indent As an attempt to offer a solution to these problems, a \textit{Multiresonant Layered Acoustic Metamaterial} (MLAM) is proposed in the present work.~Its constant thickness, thin layer-based configuration allows it to be easily manufactured by presently available well-known techniques. These could refer, for instance, to a process involving lamination and die cutting.~While multi-layer panels based on acoustic metamaterials have been proposed in the past, they typically rely on membrane-type metamaterials in which different material components are combined in the same layer, which still represents an issue in terms of the manufacturing process (see, for instance, Refs.~\cite{Yang2010,Naify2012}).~Furthermore, in order to improve its attenuating capabilities in target frequency ranges, the internal resonances of different layers of the MLAM proposed here are coupled, which translates into an extended effective attenuation band exhibiting a double-peak sound transmission loss (STL) response.~The physical background behind the coupled resonances mechanism, as well as its effects in terms of the system's dispersion characteristics, is described theoretically in \citet{Gao2020} through simple spring-mass systems, but without providing an actual practical implementation.~Here, we present a practical realization of the concept and analyse the consequences in terms of the STL response.\\
\indent In this article, we show a methodology to numerically compute the STL response of metamaterial-based panels under normal-incidence acoustic waves.~This methodology will be used to characterize the proposed MLAM attenuating capabilities, highlighting the benefits of the coupled resonances mechanism when compared to other conventional configurations.~Finally, the concept will be validated through an impedance tube measurement of a 3D-printed MLAM early prototype.

\begin{figure}
	\includegraphics{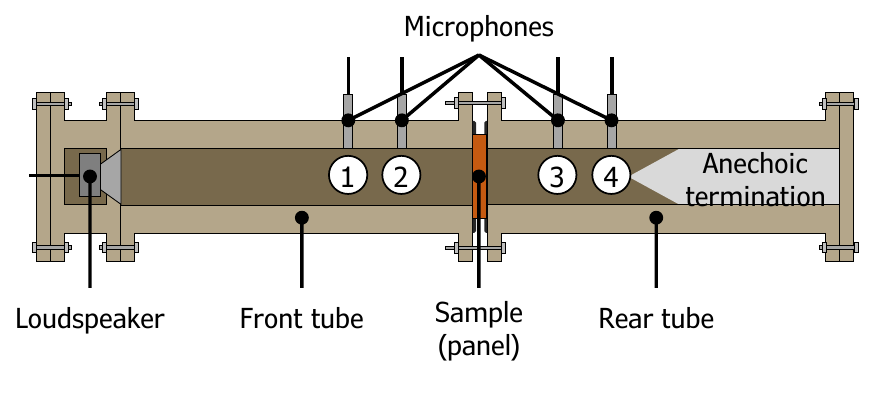}
	\caption{Experimental setting schematic representation of the two impedance tubes.~The loudspeaker generates acoustic pressure plane waves on the front tube, with a fraction being reflected back by the sample panel and the remaining being transmitted through it to the rear tube. The rear tube contains an anechoic termination to avoid spurious reflections. Four microphones measure the acoustic pressure at the designated positions, which can be used to obtain the STL of the sample panel. \label{Fig2}}
\end{figure}

\section{Materials and methods}
\label{sec:2}

\subsection{Experimental test}
\label{sec:2.1}
\indent The experimental test considered to evaluate the performance of the prototype panel consists of an impedance tube measurement of its normal-incidence STL.~This method has been used in previous works with the same purpose, to evaluate the performance of LRAM-based panels \cite{Roca2020}.~Since a detailed description of the whole methodology can be found in \citet{Roca2020}, here we will focus only on the most relevant aspects of it.\\
\indent The impedance tube setup (depicted in Fig.~\ref{Fig2}) consists of two tube parts with the same constant inner section.~On one end of the front tube, a loudspeaker (VISATON FRS 8-4~Ohm, with a nominal diameter of 8~cm, 30~W, and linear frequency response from 200 and 20000~Hz) connected to a six channel amplifier (ECLER MPA 6-80) emits a pink noise (noise source on the whole audible frequency spectrum with equal amount of energy for each octave).~The last 30~cm on the opposite end of the rear tube are filled with a polyurethane foam acting as an anechoic termination.~The sample panel is placed between the two tubes, pressed against a soft sealing material that prevents sound from leaking while not fully restricting the panel's motion on the longitudinal direction.~This particular setting deviates from the standard method described in ASTM E1050-98, where the sample typically fills the inside of the tubes' section.~However, it allows more flexibility in the sample's shape while still proving to capture local resonance mechanisms on metamaterial-based panel configurations \cite{Ho2005,Roca2020}.~Finally, two pairs of 0.5~in.~prepolarized microphones (electret condenser model GRAS 40AE) with an integrated circuit piezoelectric (ICP) preamplifier (model GRAS 26CA) are located at different tube positions.~They measure the acoustic pressure at these points, from which the STL can be computed.~The tubes' section size of 8~cm~$\times$~8~cm and the separation distance between each pair of microphones of 7.6~cm guarantee an applicability range between 200 and 2000~Hz for normal-incidence acoustic plane waves.\\
\indent The procedure to compute the STL starts with the measured pressure values at the four microphone positions, which can be written in the frequency domain as
\begin{align}
	&P_1 (\omega) = A(\omega) e^{\text{i}\kappa x_1} + B(\omega) e^{-\text{i}\kappa x_1}, \ \ x_1 = -17.6\text{ cm}; \label{eq_p1}\\
	&P_2 (\omega) = A(\omega) e^{\text{i}\kappa x_2} + B(\omega) e^{-\text{i}\kappa x_2}, \ \ x_2 = -10\text{ cm};\\
	&P_3 (\omega) = C(\omega) e^{\text{i}\kappa x_3} + D(\omega) e^{-\text{i}\kappa x_3}, \ \ x_3 = d+10\text{ cm};\\
	&P_4 (\omega) = C(\omega) e^{\text{i}\kappa x_4} + D(\omega) e^{-\text{i}\kappa x_4}, \ \ x_4 =  d+17.6\text{ cm}, \label{eq_p4}
\end{align}
where $A$ and $B$ are the complex pressure amplitudes of the incidence and reflected waves on the front tube, respectively, $C$ and $D$ are the complex pressure amplitudes of the transmitted and reflected waves on the rear tube, respectively, $d$ is the panel's thickness, and $\kappa$ is the wavenumber, defined as
\begin{equation}
    \kappa=\dfrac{\omega}{c} \label{eq_wavenumber},
\end{equation} 
with $\omega$ being the frequency and $c$ the speed of sound in air.~Each $x_k$ in the Eqs.~\eqref{eq_p1}-\eqref{eq_p4} refers to the relative position of the $k$-th microphone with respect to the panel's incidence surface.\\
\indent The anechoic termination condition allows us to neglect the effect of the reflected wave on the rear tube (i.e., $D=0$), thus the transmission coefficient $T$ of the sample panel can be obtained simply as
\begin{equation}
	T(\omega) = \dfrac{C(\omega)}{A(\omega)}, \label{eq_coeff_T}
\end{equation}
where $A$ and $C$ are computed according to:
\begin{align}
	&A(\omega) = \text{i} \dfrac{\bar{P}_1 e^{-\text{i}\kappa x_2} - \bar{P}_2 e^{-i\kappa x_1}}{2 \sin\kappa(x_2 - x_1)}, \label{eq_coeff_A}\\
	&C(\omega) = \text{i} \dfrac{\bar{P}_3 e^{-\text{i}\kappa x_3} - \bar{P}_4 e^{-i\kappa x_3}}{2 \sin\kappa(x_4 - x_3)}. \label{eq_coeff_C}
\end{align} 
with $\bar{P}_k$ being the calibrated complex Fourier transformed pressure measured for the $k$-th microphone.~The calibration process to correct the amplitude and phase of the pressure amplitudes follows the standard ASTM E1050-98 and is described also in \citet{Roca2020}.\\
\indent The STL can finally be obtained by inserting Eqs.~\eqref{eq_coeff_A} and~\eqref{eq_coeff_C} into Eq.~\eqref{eq_coeff_T} and using the formula:
\begin{equation}
	\text{STL}(\omega) = -10 \log_{10} |T|^2. \label{eq_stl}
\end{equation}

\begin{figure}
	\includegraphics{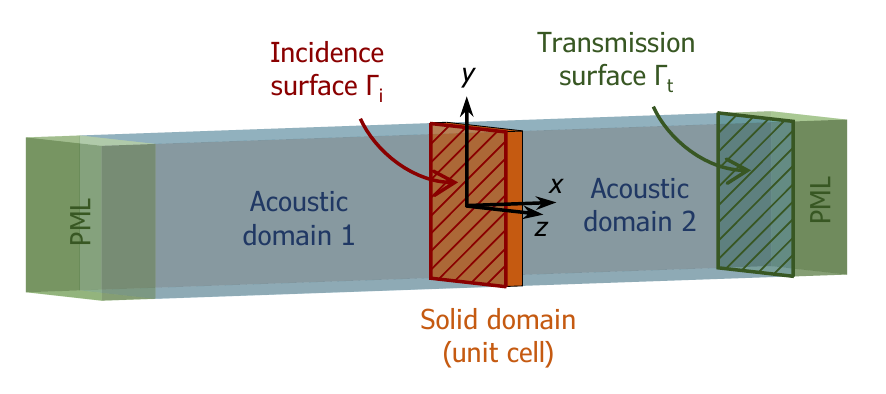}
	\caption{Schematic representation of the problem setting for numerically computing the STL of a panel with a periodically repeating unit cell.~Two perfectly matched layers (PML) are considered at the ends of the acoustic domains to simulate an infinite extension of the domain in all directions.~The STL is obtained by computing the average amplitude of the incidence and transmitted pressure fields on the designated surfaces $\Gamma_{\text{i}}$ and $\Gamma_{\text{t}}$, respectively. \label{Fig3}}
\end{figure}

\begin{figure*}
	\includegraphics{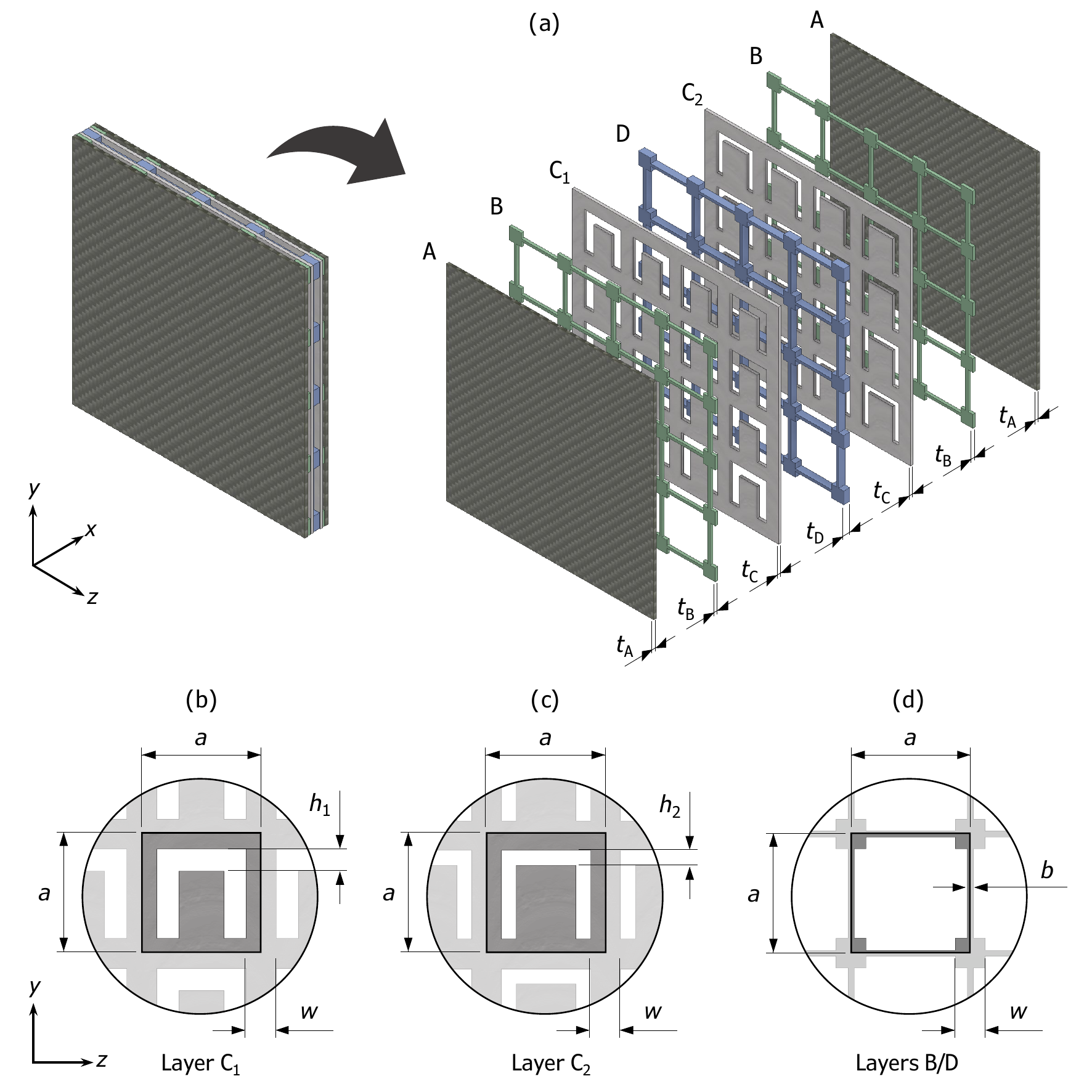}
	\caption{(a) Representation of the MLAM layered structure with an example of its different types of layers.~(b) Detailed geometrical parameters and structure of the U-shaped unit cell in the first resonating layer (C${}_1$).~(c) Detailed geometrical parameters and structure of the U-shaped unit cell in the second resonating layer (C${}_2$).~(d) Detailed geometrical parameters and structure of the unit cell in the the separating and connecting layers (B and D). \label{Fig4}}
\end{figure*}

\subsection{Numerical simulations}
\label{sec:2.2}
\indent The objective of the numerical simulations here is to compute the STL of an acoustic metamaterial panel characterized by a unit cell that infinitely repeats in the transverse directions.~To properly account for this STL parameter, the acoustic domains (filled with air, where acoustic pressure plane waves propagate), must be also infinite on each side of the solid panel.~To perform such analysis, the schematic representation in Fig.~\ref{Fig3} is considered.~The domain is split into an acoustic domain (the air), where the pressure field is the primary unknown, and a solid domain (the metamaterial panel), where the displacement field is the primary unknown.\\

\textit{(a) Acoustic domain:}\\

\indent In the acoustic domain, the pressure field is solved in terms of the frequency, $\omega$, from the Helmholtz equation
\begin{equation}
    \nabla^2 p_\text{t} + \kappa^2 p_\text{t} = 0, \label{eq_acoustic}
\end{equation}
where $\kappa$ is the wavenumber, as previously defined in Eq.~\eqref{eq_wavenumber}, and 
\begin{equation}
    p_\text{t} = p_\text{s} + p_\text{b}
\end{equation}
is the total pressure field, defined as the sum of the scattered pressure field, $p_\text{s}$ (the actual unknown), and the background pressure field, $p_\text{b}$. \\
\indent To simulate a normal-incidence acoustic plane wave, the background pressure field in the acoustic domain~1 (the incidence side) is imposed to
\begin{equation}
    p_\text{b} = p_0 e^{-\text{i}\kappa x}, \quad \forall x<0
\end{equation}
where $p_0 = 0.02$~Pa (this value corresponds to a sound pressure level of 60~dB, but is taken simply as a reference, since it does not affect the resulting STL).~In the acoustic domain~2 (transmission side), $p_\text{b} = 0$.\\
\indent At both ends of the acoustic domain (in the $x$-direction), perfectly matched layers (PMLs) are imposed to avoid wave reflections, hence simulating the medium's infinite extension.\\

\textit{(b) Solid domain:}\\

\indent In the solid domain, the displacement field, $\bs{u}$, is solved from the momentum balance equation in the frequency domain,
\begin{equation}
    \nabla\cdot\bms{\sigma} + \rho\omega^2\bs{u} = \bs{0}, \label{eq_solid}
\end{equation}
where $\rho$ is the material's density field and $\bms{\sigma}$ is the stress tensor field.~In this context, small displacements and strains can be assumed, so a linear elastic material model is considered.~In particular,
\begin{equation}
    \bms{\sigma} = \mathbb{C}:\nabla^\text{S}\bs{u},
\end{equation}
where $\nabla^\text{S}$ denotes the symmetric gradient operator and $\mathbb{C}$ is the fourth-order constitutive tensor, which can be expressed, using Voigt's notation, in terms of the material's Young's modulus, $E$, and Poisson's ratio, $\nu$, as
\begin{align}
    &\mathbb{C}^\text{Voigt} = \begin{bmatrix}
    \lambda + 2\mu & \lambda & \lambda & 0 & 0 & 0\\
    \lambda & \lambda + 2\mu & \lambda & 0 & 0 & 0\\
    \lambda & \lambda & \lambda + 2\mu & 0 & 0 & 0\\
    0 & 0 & 0 & \mu & 0 & 0\\
    0 & 0 & 0 & 0 & \mu & 0\\
    0 & 0 & 0 & 0 & 0 & \mu
    \end{bmatrix},\\
    &\text{with} \ \lambda = \dfrac{\nu E(1+\text{i}\eta)}{(1+\nu)(1-2\nu)}, \ \mu = \dfrac{E(1+\text{i}\eta)}{2(1+\nu)}, \nonumber
\end{align}
where $\eta$ is used here as an isotropic loss factor to account for damping in the system (notice that $\text{i}$ refers to the imaginary number, so the tensor $\mathbb{C}$ becomes complex valued for $\eta\neq0$).~It should be remarked that damping does not play a major role into the expected material behaviour (it is introduced simply to characterize a more realistic response), so this simple model provides just a measure of each material's dissipation capacity, regardless of the specific sources (it takes into account, for instance, visco-elastic or thermo-viscous effects).\\

\textit{(c) Acoustic-structure interface:}\\

\indent On the boundaries between the solid and acoustic domains, the effect of the pressure field on the solid domain is accounted for as an external traction force
\begin{equation}
    \bms{\sigma}\cdot\bs{n} = -p_\text{t} \bs{n}, \label{eq_boundary1}
\end{equation}
where $\bs{n}$ is the outward unit vector of the solid's boundary surface.~Additionally, compatibility of the normal component of the displacements is considered through
\begin{equation}
    \bs{n}\cdot \nabla p_\text{t} = \omega^2\rho_\text{a} \bs{u}\cdot\bs{n}, \label{eq_boundary2}
\end{equation}
where $\rho_\text{a}$ is the air's density.\\

\textit{(d) Periodic boundary conditions:}\\

\indent Given the periodic configuration of the acoustic metamaterial structure, the analysis is focused on a single unit cell while still accounting for the infinite transverse extension of the panel.~In general, this is achieved through imposing Floquet boundary conditions in the $y$ and $z$-directions, both on the solid and the acoustic domains:
\begin{align}
    & p_{\text{t}}(\bs{x}^{(+)}) = p_{\text{t}}(\bs{x}^{(-)})  e^{\text{i}\bms{\kappa}\cdot(\bs{x}^{(+)}-\bs{x}^{(-)})} \label{floquet_1}\\
    & \bs{u}(\bs{x}^{(+)}) = \bs{u}(\bs{x}^{(-)}) e^{\text{i}\bms{\kappa}\cdot(\bs{x}^{(+)}-\bs{x}^{(-)})} \label{floquet_2}
\end{align}
where $\bs{x}^{(-)}$ and $\bs{x}^{(+)}$ refer to the source and corresponding destination points, respectively, of the periodic boundaries.\\
\indent For this particular case, in which we consider waves propagating in the $x$-direction (normal-incidence, i.e. $\kappa_y = \kappa_z = 0$), conditions expressed by Eqs.~\eqref{floquet_1} and \eqref{floquet_2} turn into standard periodic boundary conditions.\\

\textit{(e) STL computation:}\\

Equations~\eqref{eq_acoustic} and \eqref{eq_solid} are solved considering a FEM discretization using the standard coupled acoustic-solid model in COMSOL.~Once the pressure and displacement fields are obtained, for a given frequency, the corresponding STL is computed as in Eq.~\eqref{eq_stl}, with the transmission coefficient, $T$, obtained through
\begin{equation}
    T = \dfrac{\int_{\Gamma_\text{t}} p_\text{t} \text{d}\Gamma}{\int_{\Gamma_\text{i}} p_\text{b} \text{d}\Gamma},
\end{equation}
where $\Gamma_\text{i}$ and $\Gamma_\text{t}$ denote the incidence and transmission surfaces, respectively, as depicted in Fig.~\ref{Fig3}.

\begin{table*}
    \centering
    \begin{tabular}{p{3.5cm}C{2.5cm}C{2.5cm}C{2.5cm}C{2cm}C{2.5cm}}
        \hline
         Material & \thead{Density\\$\rho$ (kg/m${}^3$)} & \thead{Young's modulus\\$E$ (MPa)} & \thead{Poisson's\\ratio $\nu$} & \thead{Loss\\factor $\eta$} & Layers           \\
         \hline
         Polyamide (PA)               & 1050        & 1650           & 0.4              & 0.005       & A, B             \\
         Steel                        & 7800        & 200000         & 0.3              & 0           & C${}_1$, C${}_2$ \\
         Silicone rubber              & 1050        & 0.15           & 0.47             & 0.02        & D                \\
         \hline
         \hline
         Parameter                    & MLAM        & UMLAM          & SLAM             & HP          & HP+              \\
         \hline
         $t_{\text{A}}$ (mm)          & 1 / 1       & 1 / 1          & 2.9 / 2.9        & 9.5         & 38               \\
         $t_{\text{B}}$ (mm)          & 0.5 / 0.5   & 0.5 / 4 / 0.5  & 1 / 1            & -           & -                \\
         $t_{\text{C}_1}$ (mm)        & 0.75        & 0.75           & 0.75             & -           & -                \\
         $t_{\text{C}_2}$ (mm)        & 0.75        & 0.75           & -                & -           & -                \\
         $t_{\text{D}}$ (mm)          & 4           & -              & -                & -           & -                \\
         $a$ (mm)                     & 40          & 40             & 40               & -           & -                \\
         $b$ (mm)                     & 1           & 1              & 1                & -           & -                \\
         $w$ (mm)                     & 4           & 4              & 4                & -           & -                \\
         $h_1$ (mm)                   & 8           & 8              & 6                & -           & -                \\
         $h_2$ (mm)                   & 4           & 4              & -                & -           & -                \\
         \hline
         Total thickness (mm)         & 8.5         & 8.5            & 8.55             & 9.5         & 38               \\
         \hline
         Surface density (kg/m${}^2$) & 9.96        & 9.96           & 9.95             & 9.98        & 39.9             \\
         \hline
    \end{tabular}
    \caption{Material properties and geometrical parameters considered for obtaining the STL curves in Fig.~\ref{Fig5}.~Each material is assigned to the corresponding layer types (as defined in Fig.~\ref{Fig4}) indicated in the layers column.~The definition of all the geometric parameters can also be found in Fig.~\ref{Fig4}.~The last two rows in the table provide detailed information regarding each panel configuration's total thickness and surface density, respectively.}
    \label{Tab1}
\end{table*}

\section{Proposed design}
\label{sec:3}

\begin{figure*}
	\includegraphics{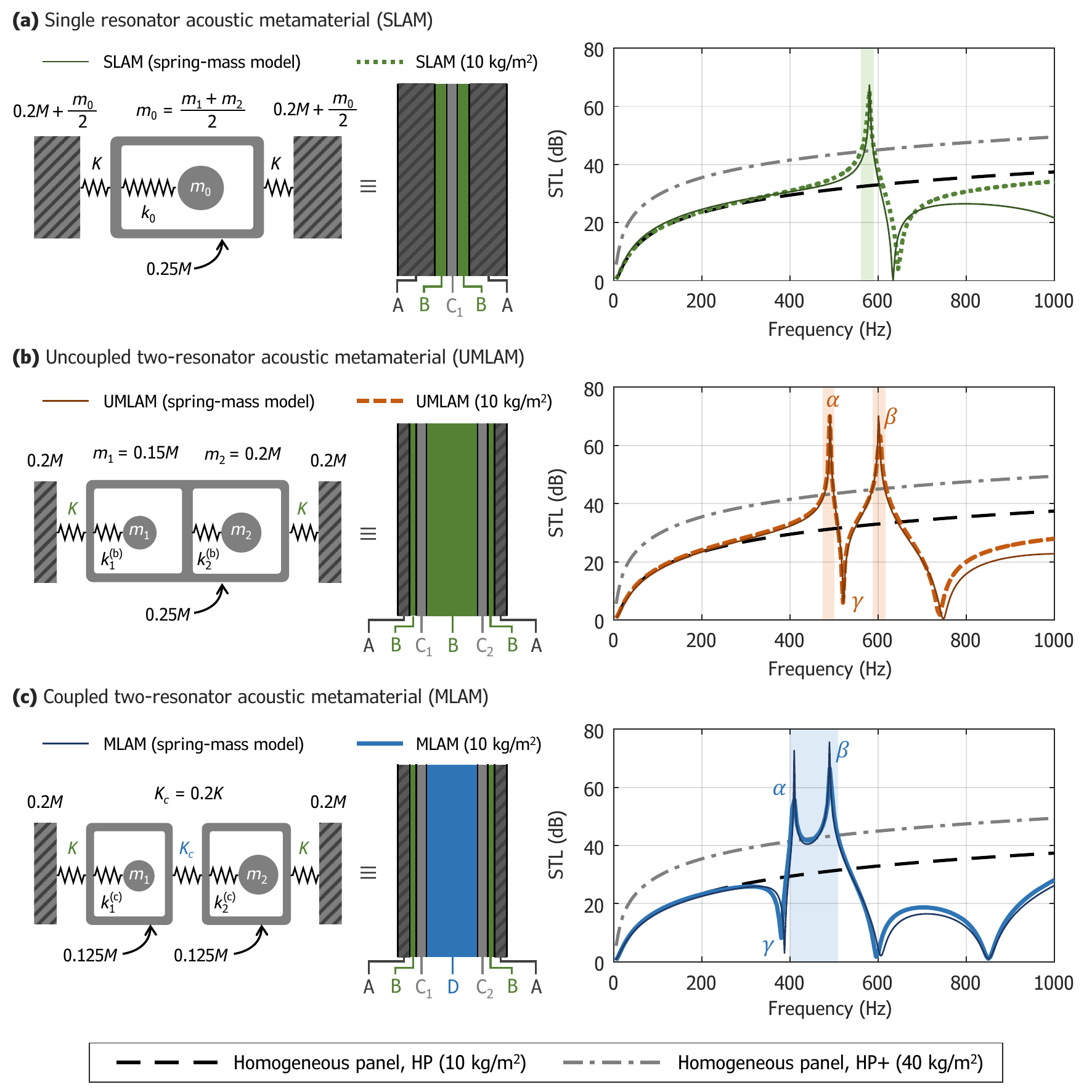}
	\caption{On the left, the three different layered-based acoustic metamaterial configurations are depicted, with their respective layer types composition, along with an equivalent spring-mass system representation.~On the right, the corresponding STL curves for each case are shown and compared to the results for homogeneous panels of the same surface density (10~kg/m${}^2$) and four times that (40~kg/m${}^2$).~The shaded areas in the plots correspond to frequency regions in which the metamaterial configurations exhibit levels of attenuation higher than those obtained with comparable homogeneous panels with four times more mass.~For reference, the thin continuous lines on the right plots represent the STL curves for each equivalent spring-mass system.\label{Fig5}}
\end{figure*}

\indent The MLAM design concept is based on a stack of \textit{constant thickness} layers of \textit{homogeneous materials}.~Each of these layers is characterized by its specific material properties, thickness and unit cell's design/topology (see Fig.~\ref{Fig4}(a)). In order to operate properly (i.e., exhibit the coupled resonances and the double-peak effects), different layers with particular purposes are required:
\begin{itemize}
    \item [a)] \textit{Resonating layers} (e.g.~layers C${}_i$ in Fig.~\ref{Fig4}(a)).~They represent the core layers of the MLAM structure, as they contain the resonating elements responsible for triggering local resonance effects, causing the STL attenuation peaks.~In this kind of layers, the unit cell's topology and dimensions play an important role in determining the location of these peaks on the frequency spectrum.~The simple ``U-shaped'' design considered for the proposed prototype (see Fig.~\ref{Fig4}(b)-(c)) is enough to guarantee a good STL response in the frequency range below 1~kHz.~In this case, the overall performance of the STL is improved when choosing denser materials for the resonating layers, since it allows them to be more compact and thin.
    \item [b)] \textit{Connecting layers} (e.g.~layer D in Fig.~\ref{Fig4}(a)).~As the name suggests, the purpose of these layers is to connect pairs of resonating layers, and so they must be stacked between them as depicted in Fig.~\ref{Fig4}(a).~Their presence in the MLAM structure is key to enable the coupling resonances effects.~In this case, their associated unit cells' pattern must be compatible with that of the resonating layers', making sure the corresponding unit cells' boundaries match (see Fig.~\ref{Fig4}(c)).~The specific dimensions and material choice for this kind of layers is also important to guarantee a good coupling effect.~This can be adjusted through the effective stiffness of the layer.~If the connecting layer is too stiff, the resonating layers become effectively uncoupled and the corresponding STL attenuation peaks wouldn't be connected (see Fig.~\ref{Fig5}).~Thus, softer (rubber-like) materials would be preferred in this case to guarantee the coupling with a more compact and thin configuration.
    \item [c)] \textit{Separating layers} (e.g.~layers B in Fig.~\ref{Fig4}(a)).~These layers are adjacent to resonating layers since their whole purpose is to enable the vibration of their resonating elements.~For this reason, their unit cells' pattern must again be compatible with that of the resonating layers.~To cause minimal interference with the overall performance of the MLAM structure, these layers should be ideally thin and stiff enough to isolate each unit cell's vibration from the others (otherwise, it can cause undesired effects on their associated local resonances due to triggering unit cell's deformation modes in the frequency range of interest).
    \item[d)] \textit{Skin layers} (e.g.~layers A in Fig.~\ref{Fig4}(a)). These layers are homogeneous solid covers (with no associated unit cell pattern), present for supporting the whole stack, and also for protection purposes.~Since they do not play a major role in the MLAM performance, they should have enough stiffness to avoid interference with the STL response.~For some applications in which the MLAM panel is attached to some structure (e.g.~a wall), the corresponding skin layer could be removed (hence attaching the separating layer directly onto the structure) while still exhibiting the relevant effects on the STL response. 
\end{itemize}

\begin{table*}
    \centering
    \begin{tabular}{p{3.5cm}C{3cm}C{3cm}C{3cm}C{3cm}}
        \hline
         Materials (3D-printing) & \thead{Density\\$\rho$ (kg/m${}^3$)} & \thead{Young's modulus\\$E$ (MPa)} & \thead{Poisson's\\ratio $\nu$} & Layers           \\
         \hline
         Polyamide (PA)               & 1010        & 1800           & 0.4              & A, B             \\
         Steel                        & 7860        & 147000         & 0.3              & C${}_1$, C${}_2$ \\
         Rubber-like                  & 1200        & 85             & 0.48             & D                \\
         \hline
    \end{tabular}
    \begin{tabular}{C{1.575cm}C{1.275cm}C{1.275cm}C{1.275cm}C{1.275cm}C{1.275cm}C{1.275cm}C{1.275cm}C{1.275cm}C{1.575cm}}
        \hline
         $t_{\text{A}}$ (mm) & $t_{\text{B}}$ (mm) & $t_{\text{C}_1}$ (mm) & $t_{\text{C}_2}$ (mm) & $t_{\text{D}}$ (mm) & $a$ (mm) & $b$ (mm) & $w$ (mm) & $h_1$ (mm) & $h_2$ (mm) \\ 
         \hline
         2 / 2 & 1 / 1 & 2 & 2 & 15 & 40 & 1 & 10 & 3 & 8 \\
         \hline
    \end{tabular}
    \caption{Material properties and geometrical parameters for the MLAM prototype used in the experiments and the corresponding numerical simulations.~Each material is assigned to the corresponding layer types (as defined in Fig.~\ref{Fig4}) indicated in the layers column.~The definition of all the geometric parameters can also be found in Fig.~\ref{Fig4}.~The material properties are obtained from the datasheets of the manufacturer.}
    \label{Tab2}
\end{table*}

\section{Results}
\label{sec:4}

\subsection{Double-peak STL response}
\label{sec:4.1}
\indent In Fig.~\ref{Fig5}, we show the STL response of three different layered-based acoustic metamaterial configurations: (a) one with a single resonating layer (SLAM), (b) another with two uncoupled resonating layers (UMLAM), and (c) the MLAM concept (i.e., with two coupled resonating layers).~For each case, Fig.~\ref{Fig5} shows the equivalent spring-mass systems: (a) a mass-in-mass configuration describing a conventional acoustic metamaterial, (b) two different inner masses inside the same outer mass (i.e., an acoustic metamaterial with two internal resonators), and (c) two different mass-in-mass units contained in the same supercell structure (i.e., two coupled acoustic metamaterial units).~The general idea behind the MLAM realization of its equivalent spring-mass system (c) is that the cells represent the corresponding resonating layers (C${}_1$ and C${}_2$), the masses at each end are assumed as the skin layers (A), and the roles of springs $K$ and $K_c$ are taken by the separating layers (B) and connecting layer (D), respectively.~Analogous considerations can be assumed for the spring-mass systems (a) and (b) with the SLAM and UMLAM configurations, respectively.\\
\indent The material properties and geometric characteristics of each panel configuration are detailed in Tab.~\ref{Tab1}.~For comparative purposes, all three acoustic metamaterial panels have almost the same total thickness and mass.~Notice that the only difference between the UMLAM and the MLAM is the material choice for the layer between the two resonating layers.~For the uncoupled configuration (UMLAM), the same polymer-like material of the separating layers (type B layers) is considered, while a softer rubber-like material (with the same density) is chosen for the MLAM case, making it a truly connecting layer (type D layer), and hence exhibiting the coupled resonances effect and the double-peak STL response.~In the SLAM configuration, the skin and separating layers' thicknesses have been increased to match the global mass and thickness of the other two configurations.~To properly analyse the STL response, the obtained curves for each case in Fig.~\ref{Fig5} are compared with the corresponding results for reference homogeneous panel (HP) configurations of the same surface density (of 10~kg/m${}^2$) and 4 times that (40~kg/m${}^2$, named HP+).~As an example, these could refer to steel panels of 1.28~mm and 5.12~mm thick, respectively, or polyamide (PA) panels of 9.5~mm and 38~mm thick (considering the densities provided in Tab.~\ref{Tab1}).\\
\indent Additionally, to support the physical insight provided by the equivalent spring-mass systems, the STL plots in Fig.~\ref{Fig5} also show their corresponding STL curves.~To obtain these results, the conditions of incident, reflected and transmitted plane waves have been imposed in terms of non-dimensional displacements and forces acting on the end masses:
\begin{align}
& \hat{u}_\text{L} = (1-R)e^{\text{i}\omega t}, & & \hat{u}_\text{R} = Te^{\text{i}\omega t};\label{eq_simple1}\\
& \hat{f}_\text{L} = \text{i}\hat{\omega}\gamma(1+R)e^{\text{i}\omega t}, & & \hat{f}_\text{R} = -\text{i}\hat{\omega}\gamma Te^{\text{i}\omega t}.\label{eq_simple2}
\end{align}
Here, the subscripts ``L'' and ``R'' refer to the ``left'' and ``right'' end masses, respectively. Variables $R$ and $T$ are the reflection and transmission coefficients, $\hat{\omega} = \omega/\Omega_0$ (with $\Omega_0^2 = K/M$), and $\gamma = \dot{m}_{\text{air}}/M\Omega_0$ (with $\dot{m}_{\text{air}} \sim \rho_{\text{air}}c_{\text{air}}$ being a reference air mass flow rate). While no straightforward (direct) association between the springs and masses can be made with the corresponding 3D realization parameters, both systems can be made equivalent (in terms of their STL response) with an appropriate choice of the variables $\gamma=0.008$ and $\Omega_0=600$~Hz, along with the total mass $M$ distribution detailed in Fig.~\ref{Fig5} (this selection of parameters comes from attempting to fit the spring-mass model results to those for its corresponding 3D implementation).~In this case, the values considered for $k_0$, $k^\text{(b)}_i$ and $k^\text{(c)}_i$ ($i = 1,2$), defined in Fig.~\ref{Fig5}, come from matching the frequencies of their corresponding resonance STL peaks.~Notice that, for the MLAM case, a lower stiffness for the connecting spring, $K_c = 0.2K$, has been required to achieve the resonance coupling effect.
\\
\indent As expected, the more conventional acoustic metamaterial configuration, i.e. consisting of a single resonator/unit cell design, produces an STL response characterized by an attenuation peak followed by a transmission dip on frequencies that, for plane waves at normal incidence, coincide with the associated bandgap limits \cite{Roca2018,Roca2019} (see Fig.~\ref{Fig5}(a)).~The idea of exploiting the local resonance effects of different resonators can be conceived as a way to obtain higher levels of attenuation in an extended frequency range.~However, the use of multiple resonators is not enough to guarantee a broader range of attenuation in the panel's STL response.~To properly exploit them, the resonators also need to be coupled, otherwise they will produce several narrowband attenuation peaks isolated from one another (i.e., with transmission dips in-between).~Through the resonators' coupling, their locally resonant bandgaps are joined into a continuous larger gap (translating into the double-peak STL response).~This can be appreciated comparing the STL responses in Figs.~\ref{Fig5}(b) and \ref{Fig5}(c).\\
\indent Notice that through the resonating layers' coupling, one effectively makes the transmission dip that typically appears \textit{after} the STL peaks produced by local resonances (see $\gamma$ in Fig.~\ref{Fig5}(b) appearing between peaks $\alpha$ and $\beta$), to occur \textit{before} the first attenuation peak (see now $\gamma$ appearing before peaks $\alpha$ and $\beta$ in Fig.~\ref{Fig5}(c)), creating an extended single, continuous attenuation band between the two attenuation peaks.~When comparing the results with those of a homogeneous panel 4 times heavier, the MLAM configuration clearly shows how we can get the same effective attenuation levels in the whole range between 400~Hz and 500~Hz.~In both other cases, the comparative STL levels are only achieved in very narrow, isolated frequency bands (of less than 30~Hz).~It is also worth noting that this enhanced STL response occurs at lower frequencies for the MLAM configuration compared to the equivalent single resonator and two uncoupled resonators cases, which is an additional advantage.\\

\subsection{Experimental validation}
\label{sec:4.2}
\begin{figure}
	\includegraphics{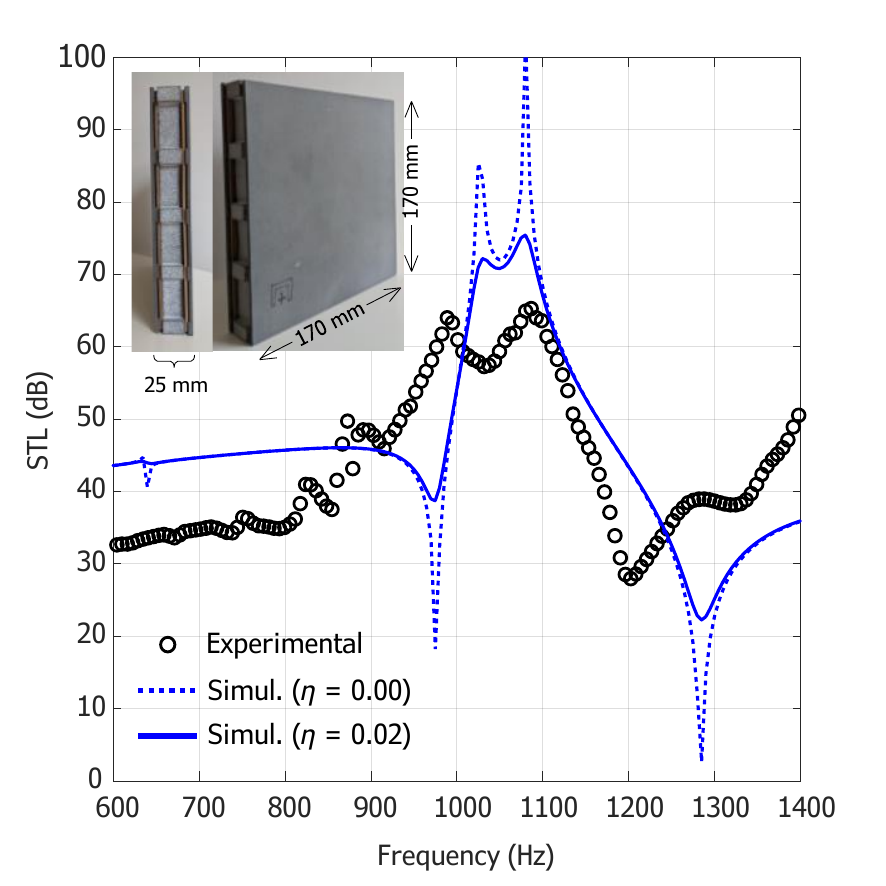}
	\caption{Sound transmission loss curve, obtained experimentally for the MLAM prototype, compared to the corresponding numerical simulation results for $\eta=0$ (no damping) and $\eta=0.02$.~The inset on the top left depicts a side view and a front view of the actual 3D-printed prototype employed with the whole panel dimensions (detailed layer composition, including geometrical parameters and materials, are given in Tab.~\ref{Tab2}).\label{Fig6}}
\end{figure}

\indent In order to validate the MLAM concept experimentally, a 3D-printed prototype of the configuration depicted in Fig.~\ref{Fig4} was tested following the procedure described in Section~\ref{sec:2.1}.~The material properties and geometrical parameters for this case are given in Tab.~\ref{Tab2}.~It should be noticed that, at present, there are limited options in terms of materials available for the 3D-printing technology and, in most cases, there are additional design constraints that need to be taken into account.~For instance, the minimum thickness of the layers.~Furthermore, since silicone rubber was not available as a material choice for the connecting layer, the closest alternative was a powder-based material that, in its final form (achieved through a Multi-Jet Fusion process), it exhibits some rubber-like behaviour.~While being certainly softer than most conventionally polymer-based materials, this material's stiffness is still 1-2 orders of magnitude higher than that of common silicone rubber materials, which required a noticeable increase of the connecting layer's thickness to compensate, and make the coupled resonances work.~Once each different layer component was 3D-printed, they were glued together forming a MLAM stack.~The tackled frequency range of attenuation for the prototype is around 1000~Hz (in contrast to the 400-500~Hz range considered in Fig.~\ref{Fig5} numerical simulations) due to the material options and geometric tolerances for the available manufacturing process.\\
\indent Four different tests were performed on the same prototype, with Fig.~\ref{Fig6} showing the average of the obtained results.~The results show more than 20~dB increase over a range of 200~Hz between two clearly defined peaks at 1000~Hz and 1100~Hz.~These experimental results are consistent with the expected STL curves obtained from the numerical simulation, especially when considering an appropriate amount of damping, with a loss factor of $\eta=0.02$ (expressly chosen to match to the experimental results).~The experiment setting does not reproduce the ideal conditions of the numerical simulations, which explains some of the differences between both curves (like, for instance, the appearance of small spurious peaks between 800 and 900~Hz).~The lower STL level achieved in the experimental test for lower frequencies (minus 10~dB compared to the simulation results) is likely due to small sound leaks, caused by a non-perfect isolation inside the tube and to the particular sample holding approach.~Finally, dimensional tolerances during the manufacturing process (and possible inaccuracies in the material properties provided by the manufacturer that have been used in the simulations) may be cause of small frequency shifts in the expected STL peaks.~Despite these small discrepancies the experimental results clearly show the expected enhanced STL response, with an extended effective attenuation band due to two joined STL peaks in the predicted frequency range.

\section{Conclusions}
\label{sec:5}
\indent Results for the MLAM design proposed show great potential for it to become a practical lightweight, sound attenuating solution at lower frequency ranges.~Its main advantage over other conventional acoustic metamaterial configurations stems from its layer-based structure, that makes it suitable for large-scale production through well-established manufacturing processes.~Additionally, by taking advantage of the coupled resonances phenomenon combined with the acoustic metamaterials' local resonance effects, it manages to produce STL responses exhibiting wider effective attenuation bands at lower frequencies, compared to equivalent single-resonator or uncoupled multi-resonator based acoustic metamaterials.\\
\indent While the proposed designs in the present work provide a first notion of what are the capabilities of the MLAM technology, along with its validation through an early 3D-printed prototype, there are still issues to tackle.~Early studies suggest that there is a compromise between extending the attenuation bandwidth (by separating the resonance frequencies) and the STL level achieved.~This affects also the coupling mechanism, which is not triggered when these resonances are separated beyond a certain value.~For the samples presented in Fig.~\ref{Fig4}, attenuation bandwidths of up to 200~Hz can still be achieved, but with significantly lower STL performances.~However, this limit is linked to material and geometrical features of the MLAM design, and a more detailed analysis would be a topic for future research.~In this regard, exploring the effects of coupling several MLAM panels or investigating the role of the layers' design and topology into optimizing the STL response could lead to more efficient performances (either by increasing the attenuation level or further extending the effective attenuation band).~Also, it would be interesting to perform experimental tests on prototypes produced with more suitable manufacturing techniques (for instance, lamination combined with die cutting) to analyze possible challenges and limitations of the technology that would be considered in future, more optimized MLAM designs.

\section*{Acknowledgements}
This research has received funding from the European Research Council (ERC) under the European Union's Horizon 2020 research and innovation program (Proof of Concept Grant call reference
ERC-2019-PoC-25-04-2019, Proposal No. 874481) through the project ``Computational design and prototyping of acoustic metamaterials for tailored insulation of noise'' (METACOUSTIC).~The authors also acknowledge the financial support received by the Spanish Ministry of Economy and Competitiveness, through the Research Grant DPI2017-85521-P for the project ``Computational Design of Acoustic and Mechanical Metamaterials'' (METAMAT) and the ``Severo Ochoa Programme for Centres of Excellence in R\&D'' (CEX2018-000797-S).~The authors would like to acknowledge Prof. Mahmoud Hussein, for the insights behind the physics of the coupled resonances mechanism.~Finally, the authors would like to gratefully acknowledge Dr.~Jordi Romeu and the LEAM group for facilitating the realization of the experimental part of this work.

\bibliographystyle{plainnat}
\bibliography{Biblio}

\end{document}